\documentclass[twocolumn,twoside,showpacs,preprintnumbers,aps,pre,floatfix]{revtex4-1}
\pdfoutput=1
\usepackage[pdftex]{graphicx} 
\usepackage{amsmath}
\usepackage{amssymb}
\usepackage{color}
\usepackage[utf8]{inputenc}
\newcommand{\Eq}[1]{Eq.~(\ref{eq:#1})}
\newcommand{\Eqs}[1]{Eqs.~(\ref{eq:#1})}
\newcommand{\eq}[1]{(\ref{eq:#1})}

\newcommand{\Fig}[1]{Fig.~\ref{fig:#1}}

\newcommand{\dd}{\ensuremath{\, \mathrm{d}}}

\begin{document}

\title{Granular Brownian Motors: role of gas anisotropy and inelasticity}
 
\author{
 Johannes Blaschke
} 
\affiliation{
  	Max-Planck Institute for Dynamics and Self-Organization (MPI DS), 
  	37077 G\"ottingen,
  	Germany
}
\affiliation{
	%Fakult\"at f\"ur Physik, Georg-August-Universit\"at G\"ottingen, 37077 G\"ottingen, Germany
	Faculty of Physics, Georg-August-University Göttingen, 37077 Göttingen, Germany
}
\author{
 J\"urgen Vollmer
} 
\affiliation{
  	Max-Planck Institute for Dynamics and Self-Organization (MPI DS), 
  	37077 G\"ottingen,
  	Germany
}
\affiliation{
	%Fakult\"at f\"ur Physik, Georg-August-Universit\"at G\"ottingen, 37077 G\"ottingen, Germany
	Faculty of Physics, Georg-August-University Göttingen, 37077 Göttingen, Germany
}

\date{\today}

\begin{abstract}

We investigate the motion of a 2D wedge-shaped object (a \emph{granular Brownian motor}), which is restricted to move along the $x$-axis and cannot rotate, as gas particles collide with it. We show that its steady-state drift, resulting from inelastic gas-motor collisions, is dramatically affected by anisotropy in the velocity distribution of the gas. We identify the dimensionless parameter providing the dependence of this drift on shape, masses, inelasticity, and anisotropy: the anisotropy leads to dramatically enhanced drift of the motor, which should easily be visible in experimental realizations.

\end{abstract}


\maketitle

\begin{figure}
\includegraphics[width=0.40\textwidth]{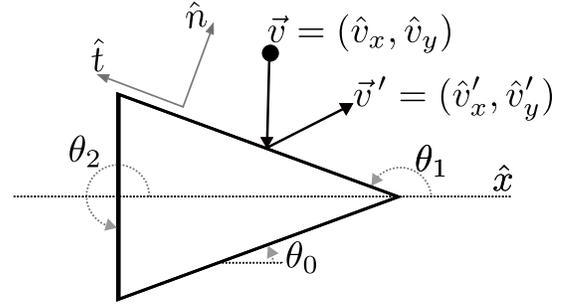}
\caption{A particle (black circle) colliding with the Brownian motor (triangular wedge with wedge angle $2 \theta_0$). The angles of the edges, $i\in\{0,1,2\}$, are measured counter-clockwise from the positive $x$-axis to the \emph{outside} of the motor, yielding $\theta_0$,  $\theta_1=\pi-\theta_0$, and $\theta_2=3\pi/2$, respectively.}

\label{fig:fig1}
\end{figure}

\emph{Introduction} --- We investigate the motion of a 2D wedge-shaped object, which we shall refer to as  the \emph{motor} (\Fig{fig1}). It cannot rotate and is restricted to move along the $x$-axis, as gas particles collide with it. When the motor experiences elastic collisions, there is a finite transient drift as the motor approaches thermal equilibrium with the gas \cite{Sporer:2008ck}. A finite steady-state motion is achieved when the gas-motor collisions are inelastic  \cite{Cleuren:2007di,Costantini:2008ur,Costantini:2010kd,Joubaud:2012df,Gnoli:2013vx}. The latter systems have consequently been called \emph{granular Brownian motors}. 

They are prototypes of systems where small particles collide with heavy objects that break reflection symmetry. Such models have been used to explore the rectification of thermal fluctuations \cite{Costantini:2008ur,Meurs:2004vu,Meurs:2005vm}, the adiabatic piston \cite{Gruber:1999wa,Piasecki:1999to}, and have lead to a novel treatment of non-equilibrium steady-states \cite{Fruleux:2012dm}. In an experimental realization \cite{Joubaud:2012df}, it was demonstrated that they even obey non-equilibrium fluctuation theorems.

So far, however, all pertinent theoretical studies are based on thermostatted gasses such that impacting particles are sampled from a Maxwellian velocity distribution. When thermostatting via stochastic forcing, this is a reasonable assumption \cite{Costantini:2010kd}. On the other hand, experimental realizations of granular gasses typically exhibit sustained heterogeneities in density and granular temperature \cite{Clewett:2012cb,Roeller:2011kq,Eshuis:2010ij,Eshuis:2005bq,Royer:2009wt}. Moreover, when shaking in the plane of observation, they exhibit noticeable anisotropy of the granular temperature \cite{Meer:2007io}. Consequentially we denote them as anisotropic gasses.

Here, we revisit the approach by which \cite{Cleuren:2007di,Meurs:2005vm,Meurs:2004vu} derived the theory for the isotropic case. Then we address the motion of the motor driven by an anisotropic gas.

%%%%%%%%%%%%%%%%%%%%%%%%%%%%%%%%%%%%%%%%%%%%%%%%%%%%%%%%%%%%%%

\emph{Gas Velocity Distribution Function (VDF)} --- Following \cite{Meer:2007io}, we model an anisotropic VDF using a squeezed Gaussian,
\begin{equation}
\phi(\hat{v}_x,\hat{v}_y)=\frac{m}{2 \pi k T}\exp{\left[-\frac{m}{2}\left(\frac{\hat{v}_x^2}{kT_x}+\frac{\hat{v}_y^2}{
kT_y}\right)\right]},
\nonumber
\end{equation}
where $m$ is the particle mass, $k$ is Boltzmann's constant, $T \equiv\left\langle m (\hat{v}_x^2+\hat{v}_y^2)\right\rangle_{\phi}/(2k)$ is the gas temperature averaged over both degrees of freedom, and $T_x$ and $T_y$ are the granular temperatures in the $\hat{x}$ and $\hat{y}$ direction, respectively. Anisotropy is quantified via the squeezing parameter,  $\alpha^2:=T_y/T_x$. $\alpha\geq 1$ as we only address vertical shaking.

Introducing dimensionless velocities, $v:=\hat{v}/\sqrt{2 k T / m}$, and requiring $\phi(\hat{v}_x,\hat{v}_y)\dd \hat{v}_x \dd \hat{v}_y = \phi_\alpha(v_x,v_y)\dd v_x \dd v_y$, reduces the VDF to,
\begin{equation}
\phi_\alpha(v_x,v_y)=\frac{2}{\pi}\frac{\alpha^2+1}{\alpha} \exp{\left[-\left(\alpha^2+1\right)v_x^2-\left(\frac{1}{\alpha^{2}}+1\right)v_y^2 \right]} ,
\label{eq:phi:squeezed}
\end{equation}
which depends on $\alpha$ only, and not on $m$, $k$, $T_x$ and $T_y$.

%%%%%%%%%%%%%%%%%%%%%%%%%%%%%%%%%%%%%%%%%%%%%%%%%%%%%%%%%%%%%%

\emph{Gas-Particle Interaction} ---  A collision event is illustrated in \Fig{fig1}. The motor has dimensionless velocity $\vec{V}=V\hat{e}_x$, and a mass $M$. Collision rules depend on which side of the motor, $i\in\{0,1,2\}$, is being impacted and on the coefficient of restitution, $r$. 

Assuming no change in the tangential component of the gas particles velocity,
\begin{subequations}
\begin{equation}
\vec{v}^\prime \cdot \hat{t}_i=\vec{v} \cdot \hat{t}_i \;,
\label{eq:model:v_perp}
\end{equation}
where $\hat{t}_i=(\cos{\theta_i},\sin{\theta_i})$ is the tangential vector to the surface being impacted. In contrast, due to restitution the reflection law for the normal direction becomes,
\begin{equation}
(\vec{V}^\prime- \vec{v}^\prime)\cdot \hat{n}_i=-r \: (\vec{V}- \vec{v})\cdot \hat{n}_i \;,
\label{eq:model:v_norm}
\end{equation} 
where $\hat{n}_i=(\sin{\theta_i},-\cos{\theta_i})$ is the normal vector. Single collisions obey conservation of momentum,
\begin{equation}
v_x^\prime +\mathcal{M} V^\prime = v_x + \mathcal{M} V \;,
\label{eq:model:p}
\end{equation}
\label{eq:model}%
\end{subequations}
where $\mathcal{M}:=M/m$ is the mass ratio. Altogether \Eqs{model} determine the change in the motor velocity,
\begin{subequations}
\begin{equation}
u_i:= V^\prime - V = \gamma_i \left(v_x-V-v_y\cot{\theta_i}\right) \;,
\label{eq:rate:single_collision}
\end{equation}
where
\begin{equation}
\gamma_i \equiv \gamma(r,\mathcal{M},\theta_i) := (1+r) \frac{\sin^2{\theta_i}}{\mathcal{M}+\sin^2{\theta_i}} \;.
\label{eq:rate:gamma}
\end{equation}
\label{eq:rate}
\end{subequations}

%%%%%%%%%%%%%%%%%%%%%%%%%%%%%%%%%%%%%%%%%%%%%%%%%%%%%%%%%%%%%%

\emph{Time Evolution of the Motor VDF} --- For independent collisions, the probability density, $P_t(V)$, of finding a motor with velocity $V$ at time $t$, follows the master equation,
\begin{eqnarray}
\partial_t P_t(V) = \int_\mathbb{R}{W(V-u;u)P_t(V-u)\mathrm{d}u} \nonumber \\
 - \int_\mathbb{R}{W(V;-u)P_t(V)\mathrm{d}u},
\label{eq:master}
\end{eqnarray}
where $W(V;u)\dd u$ is the conditional probability of a motor experiencing a collision resulting in a velocity change $V \rightarrow V+u$. It can be expressed as an integral involving four specifications: selecting only those outcomes which are (i) commensurate with single collisions (\Eqs{rate}), and (ii) collide with the outside of the motor's surface; (iii) weighting single particle collisions by the impact frequency, where the collision frequency for a stationary motor is used to non-dimensionalize time; (iv) sampling over all possible impact speeds and the motor's sides, where $w_i(\theta_0)$ is the probability of picking the side $i$ \cite{Cleuren:2007di}:
\begin{widetext}
\begin{align}
W(V;u) = \sum_{i\in \{0,1,2\}} \int_\mathbb{R} \int_\mathbb{R}  
\underbrace{ \delta \left[u - \gamma(r,\mathcal{M},\theta_i) (v_x-V-v_y \cot \theta_i)\right] }_{\mathrm{(i)}}\,
\underbrace{ \Theta [ (\vec V - \vec v) \cdot \hat n_i ] }_{\mathrm{(ii)}}\,
\underbrace{ (\vec V - \vec v) \cdot \hat n_i }_{\mathrm{(iii)}}\,
\underbrace{ \phi_\alpha(v_x,v_y) \mathrm{d}v_x \mathrm{d}v_y w_i(\theta_0) }_{\mathrm{(iv)}}
\label{eq:rate:W} 
\end{align}
\end{widetext}
Consequentially the steady-state solutions of \Eq{master} are selected by $\alpha$, $\gamma(r,\mathcal{M},\theta)$, and the wedge angle $2\theta_0$.

%%%%%%%%%%%%%%%%%%%%%%%%%%%%%%%%%%%%%%%%%%%%%%%%%%%%%%%%%%%%%%

\emph{Solutions to the Master Equation using Moment Hierarchies} ---  Given that, $\forall n \in \mathbb{N}^{+}$ and $m\leq n$ the derivatives $\partial_u^m(u^nW(V;u))$ vanish for $u \rightarrow \pm \infty$, the Kramers-Moyal \cite{Risken:vl} expansion can be applied to the moments, $M_k(t):=\langle V^k \rangle = \int_{\mathbb{R}}{V^k P(V,t)} \dd V$. Together with the jump moments, $a_n(V):=\int_{\mathbb{R}}{u^n W(V;u)\dd u}$, we arrive at an evolution equation for the moments:
\begin{equation}
\partial_t M_k(t)  = \sum_{n=1}^k {k \choose n} \langle V^{k-n} a_n(V)\rangle \;.
\label{eq:rate:KM}
\end{equation}

In order to accommodate a more general velocity distribution, we compute the jump moments by expanding them as a power series,
\begin{equation}
a_n(V)=\sum_{i=0}^\infty d_{n,i}V^i \;,
\label{eq:rate:jump_moment_series}
\end{equation}
such that \Eq{rate:KM} reduces to an infinite linear system,
\begin{equation}
\partial_t M_k(t) = \sum_{l=0}^{\infty} A_{k,l}M_l \;,
\label{eq:EOM:matrix}
\end{equation}
reminiscent of a matrix equation with matrix elements,
\begin{equation}
A_{k,l}:=\sum_{j=0}^{\min\{l,k-1\}}{k \choose k-j} d_{k-j,l-j} \;.
\label{eq:EOM:matrix:elements}
\end{equation}
%corresponding to the infinite matrix \footnote{Note, the indices start at 0.}
%\begin{equation}
%A=
%\left( \begin{array}{cccc}
%0 & 0 & 0 & \cdots \\
%d_{1,0} & d_{1,1} & d_{1,2} & \cdots \\
%d_{2,0} & 2d_{1,0}+d_{2,1} & 2d_{1,1}+d_{2,2} & \cdots \\
%d_{3,0} & 3d_{2,0}+d_{3,1} & 3d_{1,0}+3d_{2,1}+d_{3,2} & \cdots \\
%\vdots & \vdots & \vdots & \ddots \end{array} \right)
%\label{eq:EOM:matrix:A}
%\end{equation}

%%%%%%%%%%%%%%%%%%%%%%%%%%%%%%%%%%%%%%%%%%%%%%%%%%%%%%%%%%%%%%

\emph{Time-resolved motor velocity-PDF} --- In general one still can not solve the infinite matrix equation~\eq{EOM:matrix}. Hence, we truncate \Eq{rate:jump_moment_series} at order $N$, which leads to,
\begin{equation}
\partial_t M_k(t) = \sum_{l=0}^{N} A_{k,l}M_l \;.
\label{eq:EOM:matrix:N}
\end{equation}
The expansion coefficients $d_{n,i}$ in \Eq{rate:jump_moment_series} are computed using the Taylor expansion coefficients $d_{n,i}=\frac{1}{n!}a_n^{(i)}(0)$ where $a_n^{(i)}(V)$ is the $i$-th derivative of the $n$-th jump moment. In order to compute these derivatives, the delta-distribution in \Eq{rate:W} is integrated out, resulting in non-trivial integrals. As long as $V=0$, these can be evaluated using \emph{Mathematica}. The higher order derivatives of these integrals are related to each other allowing them to be computed recursively. This provides an analytical, albeit tedious, expression for \Eq{EOM:matrix:N}. 

Asymptotic analysis reveals that $d_{n,i} \sim -i^{-i/2}$ for large $i$, resulting in a combined truncation error in \Eq{EOM:matrix:N} of the order of $10^{-10}$ for $N=20$. In this work, we hence solve \Eq{EOM:matrix:N} for $N=20$ and a wedge angle $\theta_0=\pi/4$ unless stated otherwise. The initial condition will always be an ensemble where all the motors are at rest: $\vec{M}(0)=(1,0,0,\cdots)$. 

\begin{figure}
\includegraphics[width=0.49\textwidth]{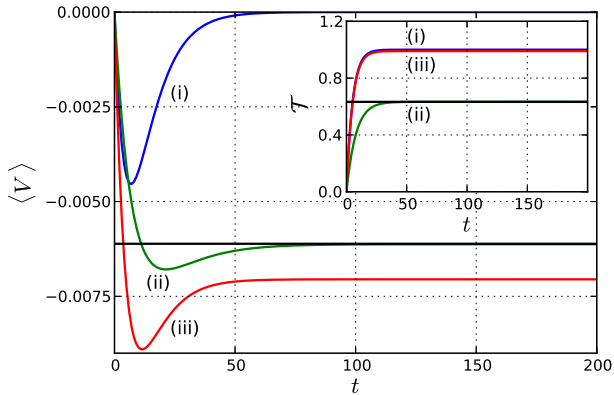} 
\caption{(Color online) Ensemble drift, $\langle V \rangle$ (main panel), and temperature, $\mathcal{T}$ (inset), against $t$ for motors with mass ratio $\mathcal{M}=10$, and $\theta_0=\pi/4$. \emph{Blue lines (i):} $r=1$ and $\alpha =1$, elastic collisions with an isotropic gas. \emph{Green lines (ii):} $r=0.3$ and $\alpha=1$, strongly inelastic collisions with an isotropic gas. The motor relaxes to the values predicted by \cite{Cleuren:2007di} (black horizontal lines). \emph{Red lines (iii):} $r=1.0$ and $\alpha=1.02$, elastic collisions with a slightly anisotropic gas.}
\label{fig:time-resolved-moments}
\end{figure}

\Fig{time-resolved-moments} illustrates typical time dependencies of the motor drift, $\langle V\rangle$, and motor temperature, $\mathcal{T}:= \mathcal{M} \left( \left<V^2\right> - \left<V\right>^2 \right)$. \emph{(i)} For elastic collisions and an isotropic gas, the ensemble undergoes a finite transient drift while it heats up to the temperature of the gas \cite{Sporer:2008ck}. Subsequently, the drift ceases. \emph{(ii)} When introducing inelastic gas-motor collisions, the steady-state acquires a finite drift velocity and a temperature significantly lower than the gas \cite{Cleuren:2007di}. \emph{(iii)} Here we note that a small amount of squeezing, $\alpha=1.02$, causes a drift similar to the drift in a system with strongly inelastic collisions. Note that this squeezing hardly affects the temperature.  

%We may compute the time dependent PDF from the higher moments in a similar fashion to \cite{Blinnikov:339055}. 
In the subsequent sections, we examine the parameter dependence of the steady-state drift, $\left<V\right>$, and motor temperature, $\mathcal{T}$, respectively.

\begin{figure*}
\includegraphics[width=0.49\textwidth]{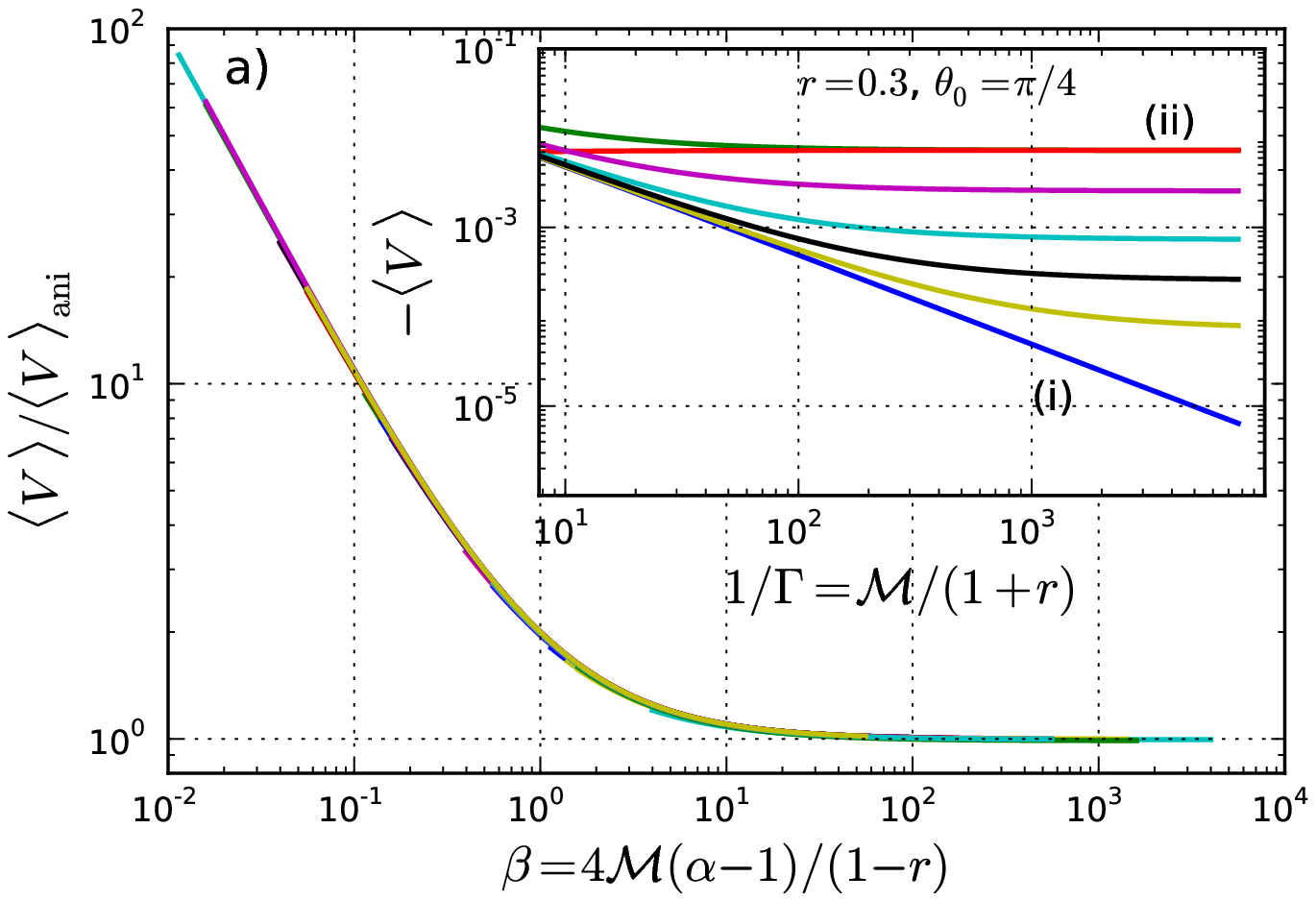} 
\includegraphics[width=0.49\textwidth]{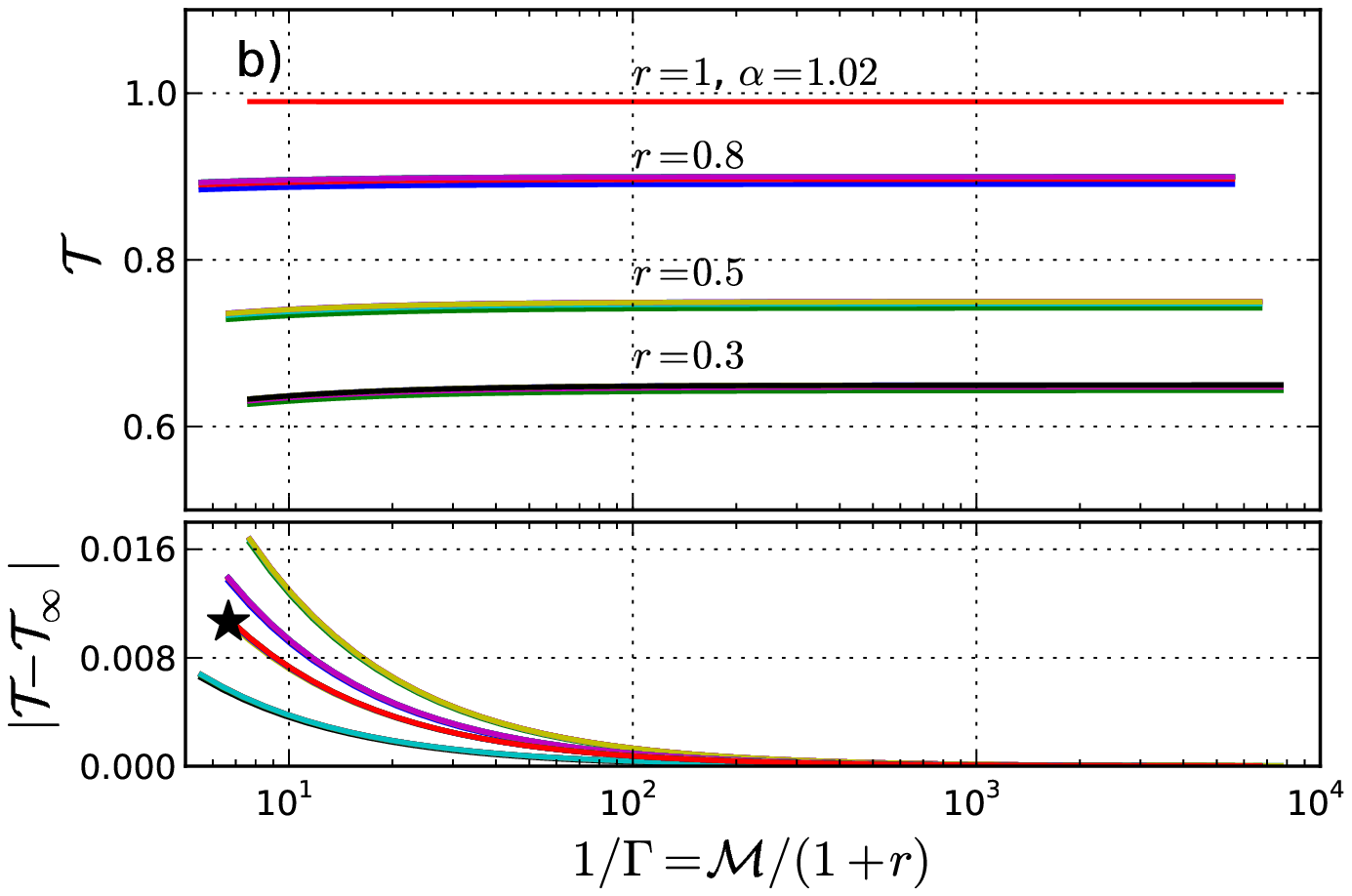}
\caption{(Color online)  Data for all combinations of $\theta_0=\pi/4$, $r\in\{0.3,0.5,0.8\}$, $\alpha\in\{1.02,1.007,1.002,1.0007,1.0002\}$; and $\theta_0=\pi/10$, $r=0.5$, $\alpha\in\{1.02,1.007,1.002,1.0007,1.0002\}$. \emph{(a)} Master plot for the motor drift where the \emph{inset} illustrates of the effect of varying~$\alpha$. The curves for inelastic collisions with an isotropic gas: $r=0.3$, $\alpha=1$ (\emph{straight blue line, (i)}); and elastic collisions with an anisotropic gas: $r=1$, $\alpha=1.02$ (\emph{straight red line, (ii)}) have been included for reference. All other curves show the drift for decreasing $\alpha\in\{1.02,1.007,1.002,1.0007,1.0002\}$ from top to bottom. \emph{(b)} \emph{(top)} The motor temperature, $\mathcal{T}$, for $\theta_0=\pi/4$, $r\in\{0.3,0.5,0.8\}$, $\alpha\in\{1.02,1.007,1.002,1.0007,1.0002\}$. \emph{(bottom)} The difference between motor temperature and the asymptotic theory. For comparison, $\theta_0=\pi/10$, $r=0.5$, $\alpha\in\{1.02,1.007,1.002,1.0007,1.0002\}$ is also shown ($\star$).}
\label{fig:thermodynamic-limit}
\end{figure*}

%%%%%%%%%%%%%%%%%%%%%%%%%%%%%%%%%%%%%%%%%%%%%%%%%%%%%%%%%%%%%%

\emph{Motor Drift} --- The inset in \Fig{thermodynamic-limit} (a) shows that for a fixed coefficient of restitution ($r=0.3$), the drift velocity initially scales as $1/\mathcal{M}$. For large $\mathcal{M}$ and $\alpha\neq 1$ it approaches a constant value depending only on $\alpha$ and $\theta_0$.  The $1/\mathcal{M}$ scaling is in agreement with the theory for the isotropic gas \cite{Cleuren:2007di}. We conclude that the drift for light motors is affected primarily by the inelastic nature of the gas-motor interactions. Here the theory for the isotropic gas is a good approximation. In contrast, massive motors are more strongly influenced by the anisotropy of the gas, no matter how slight this may be. 

In order to fully characterize this crossover, we consider the limit of a \emph{massive motor}: $\mathcal{M}\rightarrow \infty$. In this limit the $\gamma_i$ term in \Eq{rate:single_collision} simplifies,
\begin{equation}
\gamma(r,\mathcal{M},\theta)  \simeq \frac{1+r}{\mathcal{M}} \sin^2{\theta} =: \Gamma \sin^2{\theta} \;.
\label{eq:Gamma}
\end{equation}
Due to this factorization of $\sin{\theta}$ and $\Gamma$, massive motors undergoing dissipative collisions ($r<1$) behave like motors undergoing elastic collisions ($r=1$) yet with a slightly higher mass. This is in agreement with results for the granular Boltzmann equation \cite{Puglisi:2006io,Piasecki:2007hn}. Consequentially, the limit of a massive motor corresponds to the limit $\Gamma\rightarrow 0^{+}$ and is independent of restitution, $r$. 

We observe that, for small $\Gamma$, 
\begin{subequations}
\begin{align}
d_{n,i} &\sim\Gamma^{n}\\
d_{1,0} &\sim \left(\alpha -1\right) \cdot \Gamma
\end{align}
\label{eq:d:scaling}%
\end{subequations}
Hence, for isotropic gas VDFs  (where $\alpha=1$), the matrix defined by \Eq{EOM:matrix:elements} becomes upper-triangular in leading order of $\Gamma$. This corresponds to the decoupling of the time-evolution equations for the moments, as observed in \cite{Cleuren:2007di}. In contrast, for $\alpha>1$, the time evolution equations for the moments become coupled again: 
\begin{equation}
A\simeq
\left( \begin{array}{cccc}
0 & 0 & 0 & \cdots \\
d_{1,0} & d_{1,1} & d_{1,2} & \cdots \\
0 & 2d_{1,0} & 2d_{1,1}  &\cdots \\
\vdots & \vdots & \vdots & \ddots \end{array} \right)
\label{eq:lim:matrix:A}
\end{equation}

This shall be the starting point of a perturbation theory around $\left(\Gamma,\alpha\right) = (0^+,1)$. We assume that, in the limit $\Gamma\rightarrow 0^{+}$ the steady state is still largely independent of truncation size for small $(\alpha-1)$.  Hence, we find that the null space of the upper left $2\times 2$ sub-matrix of \Eq{lim:matrix:A} accurately determines the steady state drift due to anisotropy,
\begin{equation}
\left<V\right>_{\mathrm{ani}} \simeq -\frac{d_{1,0}}{d_{1,1}} \simeq \sqrt{\frac{\pi}{2}} \left(\sin{\theta_0}-1\right)(\alpha-1).
\label{eq:lim:drift}
\end{equation}
%Remarkably, only the first two expansion coefficients $d_{1,0}$ and $d_{1,1}$ are necessary for accurately determining the asymptotic drift (\Fig{thermodynamic-limit} (a)).
Note that \Eq{lim:drift} does not depend on $\mathcal{M}$. This is quite astounding since it implies that the drift \emph{velocity} of the massive motor is of the order of the gas-particle velocity (dimensionless $\left<V\right>_{\mathrm{ani}} $ is of the order $1$), even though the transferred momentum from the gas remains constant with increasing $\mathcal{M}$.

The crossover occurs when the drift for the isotropic case $\left<V\right>_{\mathrm{iso}} \simeq (1-r)\mathcal{M}^{-1} \sqrt{\pi/2} (\sin{\theta_0}-1)/4$ \cite{Cleuren:2007di} is of the same order as the drift due to anisotropy. Consequently the dimensionless number,
\begin{equation}
\beta := \frac{\left<V\right>_{\mathrm{ani}}}{\left<V\right>_\mathrm{iso}} = \frac{4\mathcal{M}(\alpha-1)}{1-r},
\label{eq:crossover}
\end{equation}
characterizes the dominant driving of the motor. For $\beta \ll 1$, the dynamics is driven by inelastic collisions ($r<1$), and for $\beta \gg 1$ the dynamics is driven by anisotropy ($\alpha>1$). 
Plotting $\langle V\rangle / \langle V \rangle_{\mathrm{ani}}$ as a function of $\beta$ provides an excellent data collapse, \Fig{thermodynamic-limit} (a).

%%%%%%%%%%%%%%%%%%%%%%%%%%%%%%%%%%%%%%%%%%%%%%%%%%%%%%%%%%%%%%

\emph{Motor Temperature} --- \Fig{thermodynamic-limit} (b) shows that the temperature is independent of $\mathcal{M}$ for $\mathcal{M}\gtrsim 10$ and it is affected by inelastic collisions more severely than by anisotropy. We now follow the perturbation theory of the previous section to determine the correction to $\mathcal{T}$ in first order of $(\alpha-1)$.

Since the motor temperature contains a coefficient of $1/\Gamma$, we must expand $A$ to second order in $\Gamma$. According to \Eqs{d:scaling} $A$ then takes the form,
\begin{equation}
A\simeq
\left( \begin{array}{ccccc}
0 & 0 & 0 & 0 & \cdots \\
d_{1,0} & d_{1,1} & d_{1,2} & d_{1,3} & \cdots \\
d_{2,0} & 2d_{1,0}+d_{2,1} & 2d_{1,1}+d_{2,2} & 2d_{1,2}+d_{2,3} & \cdots \\
0 & 3d_{2,0} & 3d_{1,1}+3d_{2,1} & 3d_{1,2}+3d_{2,2} & \cdots \\
\vdots & \vdots & \vdots & \vdots & \ddots \end{array} \right)
\label{eq:lim:matrix:A2}
\end{equation}
This results in a further increase of the coupling between the different moments. In order to reliably compute $\langle V^2 \rangle_{\mathrm{ani}}$, the null-space of at least the upper left $4\times 4$ sub-matrix of \Eq{lim:matrix:A2} must be used, yielding the asymptotic expression for the temperature,
\begin{equation}
\frac{2}{1+r} \, \mathcal{T}_{\mathrm{ani}} \simeq 1+ \left[ \frac{4-\pi}{4} (1-\sin\theta_0)^2+\sin^2\theta_0 \right](\alpha-1).
\label{eq:lim:T}
\end{equation}
The lower panel of \Fig{thermodynamic-limit} (b), shows the converge onto this asymptotic value.

%%%%%%%%%%%%%%%%%%%%%%%%%%%%%%%%%%%%%%%%%%%%%%%%%%%%%%%%%%%%%%

\emph{Conclusion} --- We have investigated the motion of a granular Brownian motor that is driven by inelastic collisions (particle-motor coefficient of restitution $r$) with an anisotropic velocity distribution (with anisotropy $\alpha-1$), modelled using a squeezed Gaussian, \Eq{phi:squeezed}.

Examining the scaling of the drift with relative motor mass, $\mathcal{M}$, we identified a crossover from the motor drift arising due to inelastic gas-motor collisions, to a setting where it arises predominantly from the anisotropy of the gas. Examining the steady-state drift of the motor in the limit of large $\mathcal{M}$, we have identified a dimensionless parameter $\beta$, \Eq{crossover} (independent of wedge angle). For $\beta\ll 1$ inelastic collisions drive the drift of the motor, and anisotropy is negligible; for $\beta\gg 1$ anisotropy dominates the drift and restitution in motor-gas collisions becomes negligible. In the latter regime we have identified a remarkably strong enhancement of the drift: it is of the order of gas particle \emph{velocity}, even in the limit of infinite motor-particle mass ratios. Is this remarkable regime accessible experimentally?

Many experiments, involving agitated granular matter, are kept in a steady state via shaking from the walls. Such systems always exhibit an anisotropic velocity distribution \cite{Meer:2007io}. Laboratory experiments can have an anisotropy of the order of $\alpha\approx 2$ \footnote{Matthias Schr{\"o}ter, private communications}, and the most conservative estimate for simulations yields $\alpha \approx 1.12$ (\cite{Meer:2007io} Fig 4, inset). Given maximally inelastic collisions ($r$ close to $0$) this amounts to $\beta \approx 0.5 \mathcal{M}$. For $\mathcal{M}>10$ typical experimental realizations therefore probe, at best, the crossover regime rather than a regime where the drift solely arises from the inelastic collisions. If one wishes to probe the latter regime, isotropy of the gas particles must be enhanced by at least two orders of magnitude for the experimental setups we are aware of.

The dramatic enhancement of the drift thus lies in an easily accessible regime, and it certainly calls for further experimental and numerical exploration.

We are grateful to P. Colberg, S. Herminghaus, R. Kapral, W. Losert, D. van der Meer, L. Rondoni, and M. Schr\"oter for enlightening discussions.

\bibliographystyle{apsrev4-1}
\bibliography{library}

\end{document}